\documentclass[aps,prl,twocolumn,tightenlines,superscriptaddress]{revtex4}
\usepackage{epsfig,rotating}
\usepackage{multirow}

\newcommand{\nuc}[2]{\hbox{$^{#1}$#2}}

\begin{document}
\title{In-beam $\gamma$-ray spectroscopy of very neutron-rich nuclei:
  Excited states in \nuc{46}{S} and \nuc{48}{Ar}}  

\author{A.\ Gade}
   \affiliation{National Superconducting Cyclotron Laboratory,
      Michigan State University, East Lansing, Michigan 48824}
   \affiliation{Department of Physics and Astronomy,
      Michigan State University, East Lansing, Michigan 48824}
\author{P.\ Adrich}
    \affiliation{National Superconducting Cyclotron Laboratory,
      Michigan State University, East Lansing, Michigan 48824}
\author{D.\ Bazin}
    \affiliation{National Superconducting Cyclotron Laboratory,
      Michigan State University, East Lansing, Michigan 48824}
\author{B.\,A.\ Brown}
    \affiliation{National Superconducting Cyclotron Laboratory,
      Michigan State University, East Lansing, Michigan 48824}
    \affiliation{Department of Physics and Astronomy,
      Michigan State University, East Lansing, Michigan 48824}
\author{J.\,M.\ Cook}
    \affiliation{National Superconducting Cyclotron Laboratory,
      Michigan State University, East Lansing, Michigan 48824}
    \affiliation{Department of Physics and Astronomy,
      Michigan State University, East Lansing, Michigan 48824}
\author{C.\ Aa.\ Diget}
    \affiliation{National Superconducting Cyclotron Laboratory,
      Michigan State University, East Lansing, Michigan 48824}
\author{T.\ Glasmacher}
    \affiliation{National Superconducting Cyclotron Laboratory,
      Michigan State University, East Lansing, Michigan 48824}
    \affiliation{Department of Physics and Astronomy,
      Michigan State University, East Lansing, Michigan 48824}
\author{S.\ McDaniel}
    \affiliation{National Superconducting Cyclotron Laboratory,
      Michigan State University, East Lansing, Michigan 48824}
    \affiliation{Department of Physics and Astronomy,
      Michigan State University, East Lansing, Michigan 48824}
\author{A.\ Ratkiewicz}
    \affiliation{National Superconducting Cyclotron Laboratory,
      Michigan State University, East Lansing, Michigan 48824}
    \affiliation{Department of Physics and Astronomy,
      Michigan State University, East Lansing, Michigan 48824}
\author{K.\ Siwek}
    \affiliation{National Superconducting Cyclotron Laboratory,
      Michigan State University, East Lansing, Michigan 48824}
    \affiliation{Department of Physics and Astronomy,
      Michigan State University, East Lansing, Michigan 48824}
\author{D.\ Weisshaar}
    \affiliation{National Superconducting Cyclotron Laboratory,
      Michigan State University, East Lansing, Michigan 48824}
\date{\today}

\begin{abstract}
We report on the first in-beam $\gamma$-ray spectroscopy study of the very
neutron-rich nucleus \nuc{46}{S}. The $N=30$ isotones \nuc{46}{S} and
\nuc{48}{Ar} were produced in a novel way in two steps that both necessarily involve
nucleon exchange and neutron pickup reactions,   
\nuc{9}{Be}(\nuc{48}{Ca},\nuc{48}{K})X followed by 
\nuc{9}{Be}(\nuc{48}{K},\nuc{48}{Ar}+$\gamma$)X at 85.7~MeV/u mid-target energy
and 
\nuc{9}{Be}(\nuc{48}{Ca},\nuc{46}{Cl})X followed by 
\nuc{9}{Be}(\nuc{46}{Cl},\nuc{46}{S}+$\gamma$)X at 87.0~MeV/u mid-target energy,
respectively. The results are compared to large-scale shell-model calculations
in the $sdpf$ shell using the SDPF-NR effective interaction and $Z$-dependent
modifications.  
\end{abstract}

\maketitle

The quest to comprehend the structure of the atomic nucleus in the
regime of large neutron excess is driving experimental and theoretical
research programs worldwide. Modifications to the familiar ordering of
single-particle orbits or new phenomena like the development of neutron halos
have been observed in experiments and their microscopic description challenges
theories that quantify the nuclear many-body system at large 
proton-neutron asymmetry. 

The experimental challenges are (i) the production of
these short-lived, radioactive nuclei and (ii) the study of their properties. Neutron-rich
nuclei lighter than calcium are efficiently produced in-flight by the
fragmentation of a stable \nuc{48}{Ca} primary beam in the collision with a
\nuc{9}{Be} target at energies exceeding 100~MeV/nucleon. The nature of this
production mechanism implies that the majority of the produced fragments has
fewer neutrons than the projectile. Reactions, however, that involve the removal of
several protons with no net loss of neutrons or additional neutrons being picked
up from the target nucleus 
proceed with comparably small cross sections (see for
example~\cite{Sol92,Pfaff95} and references within). The resulting secondary
beams of rare isotopes are typically  
available for experiments at velocities exceeding 30\% of the speed of light. A
variety of techniques have been developed to enable in-beam spectroscopy
studies of fast rare-isotope beams with intensities down to a few atoms per
second~\cite{Gade08a}. 

\begin{figure}[h]
\includegraphics[height=90mm]{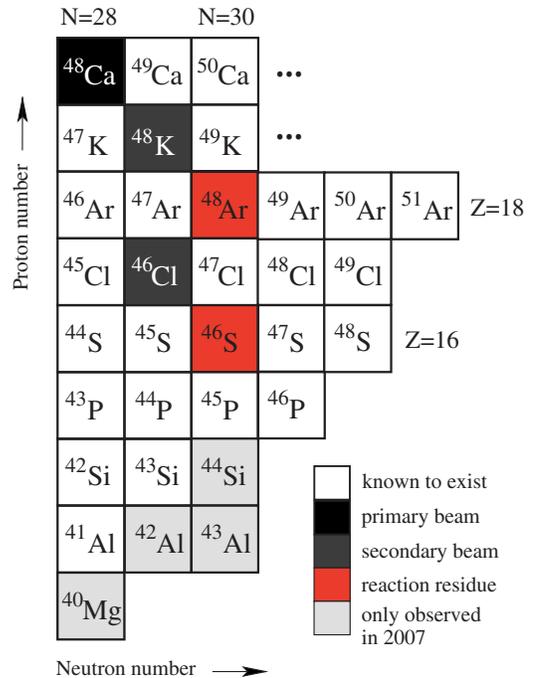}
\caption{\label{fig:chart} (Color online) Part of the nuclear chart showing the
  most neutron-rich 
  nuclei known to exist out to $Z=18$. Highlighted are  
  \nuc{48}{Ca} (primary beam), \nuc{48}{K} and \nuc{46}{Cl} (secondary beams)
  and \nuc{48}{Ar} and \nuc{46}{S} (respective final reaction residues). For comparison, the observation of \nuc{44}{Si}~\cite{Tar07},
  \nuc{40}{Mg}, \nuc{42}{Al}, and possibly \nuc{43}{Al}~\cite{Bau07} was only
  achieved in 2007. } 
\end{figure}

Here we report on the in-beam $\gamma$-ray spectroscopy of
\nuc{46}{S} 
and \nuc{48}{Ar}, each produced in a novel way in two steps that both necessarily 
involve heavy-ion induced nucleon exchange and/or neutron pickup reactions:  
\nuc{9}{Be}(\nuc{48}{Ca},\nuc{48}{K})X followed by 
\nuc{9}{Be}(\nuc{48}{K},\nuc{48}{Ar}+$\gamma$)X at 85.7~MeV/u mid-target energy
and 
\nuc{9}{Be}(\nuc{48}{Ca},\nuc{46}{Cl})X followed by 
\nuc{9}{Be}(\nuc{46}{Cl},\nuc{46}{S}+$\gamma$)X at 87.0~MeV/u mid-target energy,
respectively. \nuc{48}{Ar} and \nuc{46}{S} have neutron number $N=30$,
two neutrons more than the \nuc{48}{Ca} primary beam (see Fig.~\ref{fig:chart}). These are the
heaviest argon and sulfur isotopes studied with $\gamma$-ray
spectroscopy to date. As a result, the first excited $2^+$ state of
\nuc{46}{S} was observed for the first time. This is the first time that
\nuc{9}{Be}-induced nucleon exchange reactions at intermediate beam energies are
used to perform in-beam $\gamma$-ray spectroscopy of nuclei more neutron-rich
than the projectile.        

The region around \nuc{42}{Si} with neutron number $N=28$ has attracted much
attention in recent years. While initial one- and two-proton
knockout experiments hinted a proton sub-shell gap at
$Z=14$~\cite{Fri05}, inelastic proton scattering in the chain
of silicon isotopes~\cite{Cam06,Cam07} and ultimately the observation
of the first excited $2^+$ state of \nuc{42}{Si} at low
energy~\cite{Bastin07} revealed the breakdown of the $N=28$ shell gap
for silicon. Data on the $N=30$ isotones in this surprising 
region are scarce. Excited states in \nuc{48}{Ar} have been observed following
deep-inelastic reactions~\cite{navin} while the existence of \nuc{44}{Si} 
has only been proven recently~\cite{Tar07}. For \nuc{46}{S}, its $\beta$-decay
half-life~\cite{Gre04} had been the only observable accessible to experiments. 

Our experiments were performed at the Coupled Cyclotron Facility at NSCL on the  
campus of Michigan State University. The secondary beams of \nuc{48}{K} (pure)
and \nuc{46}{Cl} (purity exceeding 98\%) were produced from a 140~MeV/u
stable \nuc{48}{Ca} beam impinging on a 705~mg/cm$^2$ \nuc{9}{Be} production
target and separated using a 390~mg/cm$^2$ Al degrader in the A1900 fragment 
separator~\cite{a1900}. The momentum acceptance of the separator was
restricted to 0.5\% for the \nuc{48}{K} beam and to 2\% for the much less
intense \nuc{46}{Cl} beam, yielding on-target rates of
$110 \times 10^3$~particles/s and $6 \times 10^3$~particles/s, respectively.    

The \nuc{9}{Be} reaction target (376~mg/cm$^2$ thick) was surrounded by
the high-resolution $\gamma$-ray detection system SeGA, an array of
32-fold segmented HPGe 
detectors~\cite{sega}. The segmentation allows for event-by-event Doppler
reconstruction of the $\gamma$ rays emitted by the reaction
residues in flight. The emission angle entering the Doppler reconstruction is  
determined from the location of the segment with the largest  
energy deposition. Sixteen detectors were arranged in two rings 
(90$^\circ$ and 37$^\circ$ central
angles with respect to the beam axis). The 37$^\circ$ ring was
equipped with seven detectors while nine detectors were located at
90$^\circ$.  The photopeak efficiency of the array was calibrated
with standard sources and corrected for
the Lorentz boost of the $\gamma$-ray distribution emitted by nuclei
moving at 39\% of the speed of light.

\begin{figure}[h]
\includegraphics[height=50mm]{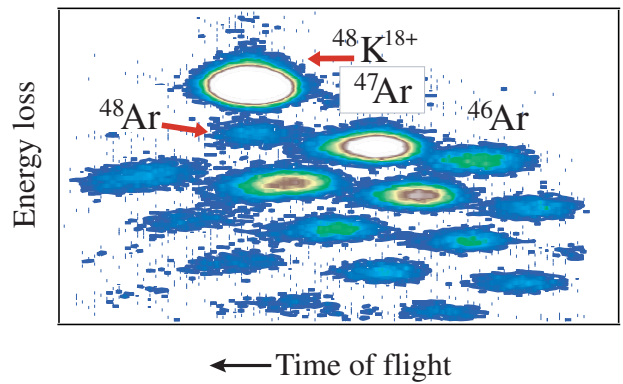}
\caption{\label{fig:pid} (Color online) Particle identification spectrum for the
  reaction 
  residues produced in \nuc{9}{Be}(\nuc{48}{K},\nuc{48}{Ar})X at
  85.7~MeV/u mid-target energy. The energy loss measured in the S800
  ionization chamber is plotted versus the ion's time of flight. \nuc{48}{Ar}
  can be 
  unambiguously separated from the projectile-like fragmentation residues
  produced in the reaction \nuc{48}{K}+\nuc{9}{Be}. }
\end{figure}

\begin{figure}[h]
\includegraphics[height=75mm]{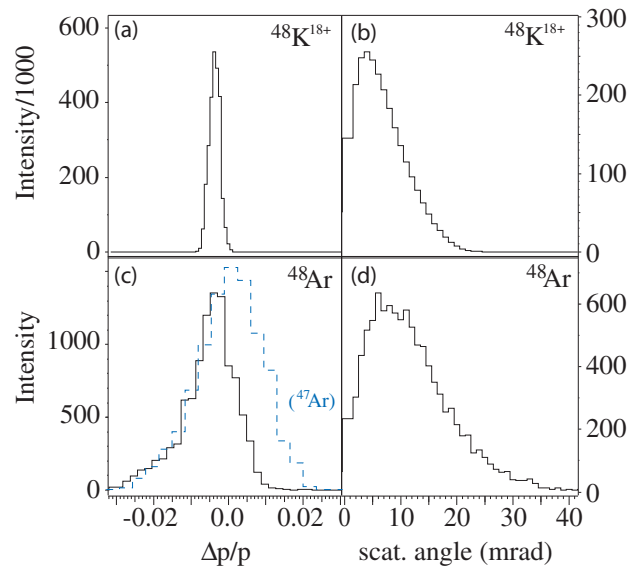}
\caption{\label{fig:all} (Color online) Parallel momentum distribution of the
  \nuc{48}{K}$^{+18}$ ions (a) and the reaction product \nuc{48}{Ar} (c) 
  relative to 18.353~GeV/c. (b) and (d) show the reconstructed scattering
  angle. The \nuc{48}{K}$^{+18}$ ions were not subject to a nuclear reaction and
  the spectra show the effect of the \nuc{9}{Be}-induced nucleon
  exchange reaction relative to the unreacted \nuc{48}{K}$^{+18}$ ions passing
  through the target. The measured longitudinal momentum 
  distribution of \nuc{47}{Ar} produced in the one-proton knockout from
  \nuc{48}{K} under identical conditions is overlayed (blue dashed line).}
\end{figure}

The particle identification was performed event by event with the
focal-plane detection system of the large-acceptance S800
spectrograph~\cite{s800}. The energy loss measured with the S800 ionization
chamber and time-of-flight information taken between plastic scintillators --
corrected for the angle and momentum of each ion -- were used to unambiguously
identify the reaction residues emerging from the target. The
particle-identification spectrum for \nuc{48}{Ar} is shown in Fig.~\ref{fig:pid}
as an example.        
The most intense constituent in the spectrum is the
\nuc{48}{K}$^{18+}$ charge state formed by the electron pickup of the initially
fully-stripped projectiles in the \nuc{9}{Be}
target. Fig.~\ref{fig:all}(a) and (b) show the longitudinal momentum
and the scattering angle distributions of the \nuc{48}{K}$^{18+}$ ions passing
through the reaction target. The parallel momentum and scattering-angle
distributions of \nuc{48}{Ar} produced in the reaction 
\nuc{9}{Be}(\nuc{48}{K},\nuc{48}{Ar})X are displayed in Fig.~\ref{fig:all}(c)
and (d). The broadening of the momentum distribution and the shift in the
maximum of the scattering angle spectrum compared to the \nuc{48}{K}$^{18+}$
ions that were not subject of a nuclear reaction (beyond scattering) are
clearly visible. However, compared to nucleon removal reactions, the momentum
distribution of the nucleon-exchange product is narrow, consistent with the
observations by Souliotis {\it et al.}~\cite{Sol92} (for comparison, the
measured parallel momentum distribution of \nuc{47}{Ar} produced in the one-proton
knockout from \nuc{48}{K} under identical conditions -- target, projectile
end momentum bite -- is overlayed).                     

Inclusive cross sections of $\sigma=0.13(1)$~mb and $\sigma=0.057(6)$~mb for the
\nuc{9}{Be}(\nuc{48}{K},\nuc{48}{Ar})X and
\nuc{9}{Be}(\nuc{46}{Cl},\nuc{46}{S})X reactions were derived from the yields of
\nuc{48}{Ar} and \nuc{46}{S} divided by the number of incoming
\nuc{48}{K} and \nuc{46}{Cl} projectiles, respectively, relative to the number
density of the reaction target. For each measurement, the normalization of the
incoming beam rate was evaluated frequently and a systematic uncertainty of 4\%
was deduced and added in quadrature to the statistical uncertainty.     

\begin{figure}[h]
\includegraphics[height=87mm]{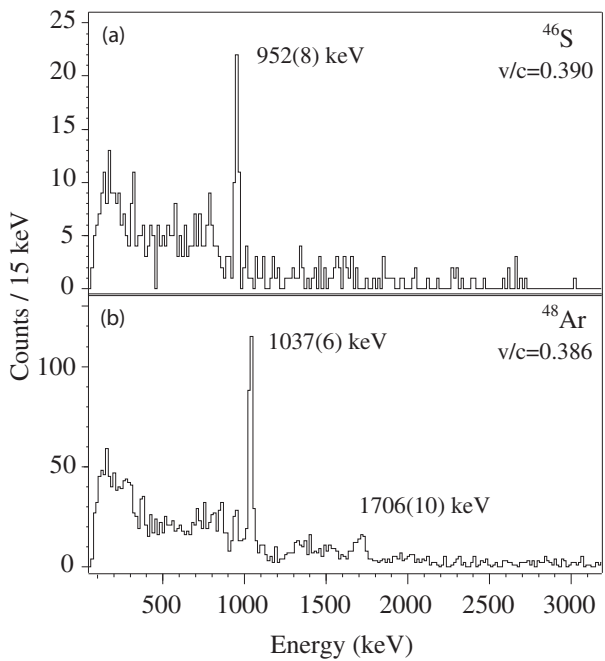}
\caption{\label{fig:gamma} Event-by-event Doppler reconstructed $\gamma$-ray
  spectra detected in coincidence with \nuc{46}{S} and \nuc{48}{Ar}. The
  952(8)~keV peak is attributed to the de-excitation of the $2^+_1$ state in
  \nuc{46}{S}. There is no evidence for peaks other than this transition. The
  two peaks observed at 1037~keV and 1706~keV for \nuc{48}{Ar} are attributed to
  the $2^+_1 \rightarrow 0^+_1$ and $4^+_1 \rightarrow 2^+_1$ transitions,
  respectively, in agreement with a previous measurement~\cite{navin}.  }
\end{figure}

The $\gamma$-ray spectra observed in coincidence with \nuc{46}{S} and
\nuc{48}{Ar} nuclei -- event by event Doppler reconstructed -- are displayed
in Fig.~\ref{fig:gamma}. The $\gamma$-ray transition at 952(8)~keV in
\nuc{46}{S} is attributed to the decay of the $2^+_1$ state to the $0^+$ ground
state. This constitutes the first observation of an excited state in this
nucleus. Gamma-ray transitions at 1037(6)~keV and 1706(10)~keV were
observed in coincidence with \nuc{48}{Ar} and assigned to the $2^+_1 \rightarrow
0^+_1$ and $4^+_1 \rightarrow 2^+_1$ transitions, respectively, in agreement
with the results of~\cite{navin}. Populations of 39(8)\% and 34(5)\% for the
$2^+$ and $4^+$ states in \nuc{48}{Ar}, respectively, were deduced from the
efficiency-corrected peak areas, while the remainder is assumed to populate the
ground state. In \nuc{46}{S}, 63(12)\% of the reactions  populate the 2$^+_1$
state. Within our limited statistics, there is no evidence for other
$\gamma$-ray transitions in the spectrum of \nuc{46}{S}.

The energy of the first $2^+$ state of \nuc{46}{S} we report
here constitutes the first measurement of an excited state in a sulfur isotope
more neutron-rich than \nuc{44}{S}$_{28}$, whose measured collectivity~\cite{Gla97}
ultimately proved changes to
the nuclear structure at neutron number $N=28$. The consistent description of
the onset of collectivity at $N=28$ in the isotopic chains of sulfur $(Z=16)$ and silicon
$(Z=14)$ has been a formidable challenge for shell-model calculations
and was achieved only recently by Nowacki and Poves in devising two effective interactions, one for $Z
\leq 14$ and one for $Z > 14$~\cite{np}. 

It is now interesting to track
the evolution of the $2^+_1$ energies beyond the eroded $N=28$ magic number in
the chains of silicon, sulfur and argon isotopes to probe the dependence of the
structure at $N=30$ on the monopole-shift and pairing modifications that were
necessary to describe the silicon isotopes within the shell
model. 

\begin{figure}[h]
\includegraphics[height=63mm]{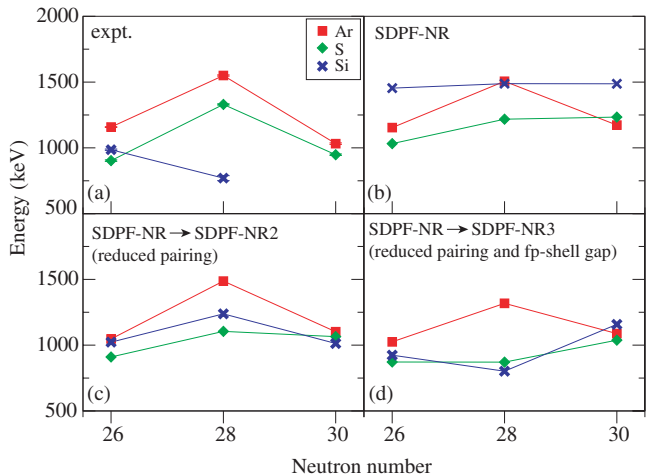}
\caption{\label{fig:energy} (Color online) Measured energies of the first $2^+$
  states in silicon, sulfur and argon isotopes with neutron numbers $N=26,~28$
  and 30 (a) compared to shell-model calculations. The calculation in (b) uses
  the SDPF-NR effective interaction~\cite{num}. The calculations for the silicon
  isotopes in (c) use the SDPF-NR2 interaction which is derived by reducing
  the $ pf $ shell $ J=0^+ $ matrix element by 0.85. The results for sulfur and
  argon in (c) are obtained from a linear interpolation in terms of $Z$ between SDPF-NR
  ($Z=20$) and SDPF-NR2 ($Z=14$). The calculations for the silicon isotopes in (d)
  use the interaction SDPF-NR3 derived by additionally lowering the neutron
  $ p_{3/2}$ and $p_{1/2}$ orbitals by 1~MeV. The results for sulfur and argon in
  (d) are obtained from a linear interpolation in terms of $Z$ between the
  SDPF-NR  ($Z=20$) and the SDPF-NR3 ($Z=14$) interactions.         }
\end{figure}

To study this systematically, shell-model configuration-interaction calculations
were carried out in the model space of the $  sd  $ shell for protons and the $
pf  $ shell for neutrons starting from the SDFP-NR interaction \cite{num} and introducing
a $Z$-dependent, linear interpolation for pairing and monopole
modifications. The experimental energies of the 2$^{ + }_1$ states 
(Fig.~\ref{fig:energy}(a)) are compared with those obtained with the SDPF-NR
Hamiltonian \cite{num} 
(Fig.~\ref{fig:energy}(b)). Experiment and theory differ in several ways, in particular
for silicon isotopes where the calculated energies are about 400~keV too high
for \nuc{40}{Si} (and for \nuc{36,38}{Si}, not shown) and 700~keV too high for
\nuc{42}{Si}.

Our first modification to SDPF-NR is a reduction of the $ fp $ shell $J=0^+$ matrix element
by 0.85 for $Z=14$. This brings the energy of the 2$^{ + }$ state
in $^{40}$Si (and $^{36,38}$Si, not shown) into better agreement with
experiment. The likely reason for this reduction, as discussed in \cite{np}, are
the reduced proton $  2p-2h  $ core-polarization contributions to the effective
interaction for silicon compared to calcium, attributed to the larger shell
gap for the orbits involved in the case of silicon (i.e., $  d_{5/2}  $ to $
f_{7/2}  $) compared to that of calcium (i.e., $  d_{3/2}  $ to $  f_{7/2}
$). These core-polarization contributions are likely the reason for the need of
two different effective interactions in this region~\cite{np}. The results for
sulfur and argon in Fig.~\ref{fig:energy}(c) were obtained with a 
linear interpolation in terms of $  Z  $ between SDPF-NR (for calcium, $  Z=20
$) to SDPF-NR2 (for silicon, $  Z=14  $). A similar argument was used to explain
the reduction of the $  sd  $ shell 
pairing in the carbon isotopes  relative to the oxygen isotopes \cite{car}.

Our second modification is to reduce the gap between the
neutron $  f_{7/2}  $ and $  p_{3/2},~p_{1/2}  $ orbitals by 1.0 MeV to obtain the
interaction SDPF-NR3 for the silicon isotopes. The results obtained with a
linear interpolation in terms of $  Z  $ between SDPF-NR (for calcium) to
SDPF-NR3 (for silicon) are shown in Fig.~\ref{fig:energy}(d). A possible reason
for the reduction of the shell gap is the lowering of single-particle energies
of low-$\ell$ orbitals ($\ell=1$) relative to those of high $\ell$ orbitals
($\ell=3$) when the energies become small as one approaches the neutron drip
line $-$ see Fig.~4 in \cite{ham}. 

The SDPF-NR3 results for 2$^{ + }_1$ energies of silicon are in good agreement
with experiment while the 2$^{ + }_1$ energies of \nuc{42}{S} and
\nuc{44}{S} obtained with the interpolated interactions are not. One expects
changes in the interaction, but it seems reasonable 
that they should be smooth as a function of proton and neutron number. In
contrast, a linear interpolation between calcium and silicon does not work $-$
and, in accordance with the proposal of \cite{np}, the interaction appears to
suddenly change at $  Z=14 $ also in our approach. The reason for this needs to be understood in
terms of a fundamental derivation of the effective interaction for this mass
region that takes the effects discussed above into account.

For the $  N=30  $ isotones studied experimentally in this work, the original
SDPF-NR interaction overestimates the 2$^{ + }_1$ energies of \nuc{48}{Ar} and
\nuc{46}{S} by 136~keV and 282~keV, respectively. The modifications
required for silicon (and implemented for sulfur and argon in terms of an
interpolated interaction), only result in a moderate
lowering of the calculated 2$^{ +}_1$ energies of the $ N=30 $ isotones,
nevertheless 
reducing the deviation between experiment and calculation SDPF-NR3 to 51~keV and
86~keV for \nuc{48}{Ar} and \nuc{46}{S}, respectively. This shows that the $
N=30$ isotones are much less impacted by the mechanisms driving the nuclear
structure at $ N=28 $.

In summary, we used the \nuc{9}{Be}-induced nucleon-exchange reactions
\nuc{9}{Be}(\nuc{48}{K},\nuc{48}{Ar}+$\gamma$)X and
\nuc{9}{Be}(\nuc{46}{Cl},\nuc{46}{S}+$\gamma$)X at above 85~MeV/nucleon
mid-target energy for the first time to perform in-beam $\gamma$-ray
spectroscopy of the $N=30$ isotones \nuc{48}{Ar} and \nuc{46}{S}.  These
heavy-ion induced nucleon-exchange reactions can lead to 
very exotic nuclei with more neutrons than the projectile beam and thus may be
considered a novel approach to reach closer toward the neutron
drip line with $\gamma$-ray spectroscopy. \nuc{48}{Ar} and \nuc{46}{S} are the
heaviest nuclei of their respective isotopic chains for which $\gamma$-ray
transitions have been measured; the  $2^+_1$ state of \nuc{46}{S} was
established for the first time. The evolution of the $2^+_1$ energies for
argon, sulfur and silicon isotopes with neutron numbers $N=26,~28$ and 30 is
tracked in comparison to large-scale shell-model calculations using the
SDPF-NR effective interaction and $Z$-dependent interpolations between the
original effective interaction and modified versions. Our studies revealed that
the description of the $N=30$ isotones improved, but at much reduced
sensitivity, when applying the monopole-shift and pairing
corrections required to describe the surprising nuclear structure at $N=28$. In 
accordance with the work by Nowacki and Poves -- the silicon isotopes emerge as 
key nuclei with a sudden change occurring in the effective interaction.

\begin{acknowledgments}
This work was supported by
the National Science Foundation under Grants No. PHY-0606007 and
PHY-0758099.
\end{acknowledgments}

\end{document}